\documentclass[twocolumn]{aastex631}

\accepted{\today}

\shorttitle{Lopsided Satellite Distributions}
\shortauthors{Samuels \& Brainerd}

\graphicspath{{./}{figures/}}

\begin{document}

\title{Lopsided Satellite Distributions around Isolated Host Galaxies in a $\Lambda$CDM Universe }

\correspondingauthor{Adam Samuels}
\email{asamuels@bu.edu}

\author[0000-0003-1895-2934]{Adam Samuels}
\affiliation{Boston University, Department of Astronomy \& Institute for Astrophysical Research \\
725 Commonwealth Ave, Boston, MA 02215, USA}

\author[0000-0001-7917-7623]{Tereasa G. Brainerd}
\affiliation{Boston University, Department of Astronomy \& Institute for Astrophysical Research \\
725 Commonwealth Ave, Boston, MA 02215, USA}

\begin{abstract}
A recent observational study found that the projected spatial distributions of the satellites of bright, isolated host galaxies tend to be lopsided with respect to the locations of the hosts. Here, we examine the spatial distributions of the satellites of a large number of bright, isolated host galaxies that were obtained from mock redshift surveys of a $\Lambda$CDM simulation. Host galaxies and their satellites were identified using selection criteria that are identical to those used in the observational study, allowing a direct comparison of the results for the simulated and observed systems. To characterize the spatial distribution of the satellites, we adopt two statistics: [1] the pairwise clustering of the satellites and [2] the Mean Resultant Length. In agreement with the observational study, we find a strong tendency for satellites in the simulation to be located on the same side of their host, and the signal is most pronounced for the satellites of blue hosts. These lopsided satellite distributions are not solely attributable to incompleteness of the observed satellite catalog or the presence of objects that have been falsely identified as satellites.  In addition, satellites that joined their hosts' halos in the distant past ($\gtrsim 8$~Gyr) show a pronounced lopsidedness in their spatial distributions and, therefore, the lopsidedness is not solely attributable to late-time accretion of satellites.
 

\end{abstract}


\section{Introduction} \label{sec:intro}

In the Cold Dark Matter (CDM) model, non-linear structure in the universe grows hierarchically (e.g., \citealt{1985ApJ...292..371D}) and, at the present day, the dark matter halos that surround galaxies, groups of galaxies, and clusters of galaxies are triaxial in shape (e.g., \citealt{2002ApJ...574..538J}).  In order to test these robust theoretical predictions in the observed universe, some type of luminous tracer of infall history and halo shape is needed.  In the case of bright, isolated galaxies that host nearby, faint satellite galaxies, the spatial distribution of the satellites may provide insight into the shapes and formation histories of the host galaxies' dark matter halos.  For example, if the satellites arrived within their hosts' halos in the far distant past, they could be expected to be a virialized population that is both symmetrically distributed about the hosts and directly traces the shapes of the hosts' dark matter halos.  If, instead, the satellites arrived within their hosts' halos relatively recently, they are less likely to be a virialized population and their spatial distributions may be neither symmetrical about the hosts nor indicative of the shapes of the hosts' halos.

Previous studies have shown that satellites around pairs of host galaxies in the observed universe are not, in fact, symmetrically distributed about each host. \citet{Libeskind_2016} showed that satellites around pairs of host galaxies in the Sloan Digital Sky Survey (SDSS;  \citealt{1996AJ....111.1748F,2001AJ....122.2129H,2002AJ....124.1810S,2000AJ....120.1579Y}) tend to occupy the space between the hosts rather than being spherically or axis-symmetrically distributed about each host. Locally, 20 of M31's 27 satellites are located between M31 and the Milky Way \citep{Conn_2013}. Lopsided satellite distributions have also been found around pairs of host galaxies in $\Lambda$ Cold Dark Matter ($\Lambda$CDM) simulations \citep{2017ApJ...850..132P,2019MNRAS.488.3100G}, and the lopsidedness has been attributed to satellites on their first orbit around their host \citep{2019MNRAS.488.3100G}. 

Due to the gravitational potential of a massive neighboring galaxy, lopsided satellite distributions between pairs of host galaxies are less surprising than might be expected for the satellites of isolated host galaxies.   A recent observational study by the authors (\citealt{Brainerd_2020}, hereafter BS20), however, found significantly lopsided satellite distributions around isolated hosts in the NASA-Sloan Atlas (NSA). BS20 computed the pairwise clustering of the satellites and found a strong tendency for pairs of satellites to be located on the same side of their host, especially in systems with blue hosts and systems with two or three satellites.

Lopsided satellite distributions have also been found around isolated hosts in $\Lambda$CDM simulations. \citet{2021ApJ...914...78W}, hereafter W21, performed an analysis of the projected spatial distribution of the satellites of isolated hosts in the Illustris-TNG300 simulation \citep{2018MNRAS.480.5113M,2018MNRAS.477.1206N,2018MNRAS.475..624N,2018MNRAS.475..648P,2018MNRAS.475..676S}. In general agreement with BS20, W21 found significant lopsidedness, especially in systems with low-mass blue hosts. However, the W21 satellite sample was selected in 3D space, with satellites being defined as all other galaxies that are located within the hosts' dark matter halo. As a result, the properties of the W21 host-satellite sample differ considerably from those in BS20.  For example, in the W21 sample, the mean number of satellites per host is a factor of $\sim 2.2$ larger than it is in BS20.  In addition, the W21 sample has many systems for which the number of satellites per host exceeds a total of 100, whereas the maximum number of satellites per system in BS20 is only 27.
 Furthermore, W21 restricted the stellar masses of their hosts to be $> 10^{11} h^{-1} M_\odot$, resulting in the majority of their host-satellite systems being more massive than those in BS20.

In this work, we directly compare the spatial distributions of satellite galaxies in the observed universe to the spatial distributions of satellite galaxies in a $\Lambda$CDM universe.  To do this, we obtain host and satellite galaxies from the $\Lambda$CDM universe using redshift space selection criteria (rather than 3D selection criteria) . To obtain the $\Lambda$CDM host-satellite sample, we employ two mock, all-sky galaxy redshift surveys based on the Millennium simulation \citep{2005Natur.435..629S}.  These mock redshift surveys allow us to select our sample from the $\Lambda$CDM universe in the same way we selected the BS20 sample from the NSA. 

In addition to allowing us to directly compare $\Lambda$CDM predictions to our previous observational results, selecting the $\Lambda$CDM host-satellite sample using redshift space criteria allows us to address potential sources of uncertainty in the observational sample that could result in a spurious detection of lopsidedness. These include candidate satellites that are within the on-sky search radius, but which lack spectroscopic redshifts (i.e., $\sim 10\%$ of galaxies in the SDSS with $r < 17.77$ lack spectroscopy due to either fiber positioning constraints or having a surface brightness that was too low for target selection).  In addition, redshift space selection of host-satellite systems always results in some fraction of the satellites being ``interlopers''. These are galaxies that pass the satellite selection criteria but are not physically associated with the host galaxies (i.e., interlopers are objects that are falsely identified as satellites). Further, unknown infall histories for observed satellite galaxies make it impossible to determine when these objects entered their hosts' dark matter halos, and it is possible that the lopsidedness of the satellite distribution is largely attributable to objects that joined their host galaxies only recently.  Here, we can directly address these issues using our simulated host-satellite sample since the 3D distances of all galaxies are known (i.e., we can determine which objects are legitimate satellites and which are interlopers), and the redshifts at which legitimate satellites joined their host galaxies' halos are also known.  This allows us to investigate the degree to which the lopsided spatial distributions of the satellites in BS20 is reproduced by a $\Lambda$CDM universe, whether the lopsidedness found by BS20 is spurious (i.e., caused by legitimate satellites that are ``missing'' from the sample due to spectroscopic incompleteness, or caused by false satellites that are located far from the host galaxies in 3D), and whether more recently accreted satellites show greater lopsidedness in their spatial distribution than do satellites that entered their hosts' halos in the distant past.

The paper is organized as follows.  In \S2 we discuss the way in which the host-satellite sample was obtained from the mock redshift surveys and we compare the properties of the simulated sample to the observational sample in BS20.  In \S3 we outline the methods we use to characterize the lopsidedness of the satellite distribution, we compare results for the simulated sample to results obtained for the BS20 observational sample, and we investigate the effects of interlopers and satellite infall redshift on the lopsidedness.  In \S4 we present a summary and discussion of the results, and in \S5 we state the major conclusions of our work.
Throughout, we adopt values of the fundamental cosmological parameters obtained by \citet{2014A&A...571A..16P}: $H_0 = 67.3$~km~s$^{-1}$~Mpc$^{-1}$, $\Omega_{m} = 0.315$, and $\Omega_\Lambda = 0.685$. 

\section{Host-satellite Sample} \label{sec:Host-satellite_Sample}

Host galaxies and their satellites were obtained from two mock all-sky galaxy redshift surveys that were created by \citet{2015MNRAS.451.2663H}. Luminous galaxies were produced with the L-Galaxies semi-analytic galaxy formation model, implemented on the Millennium dark matter simulation, and updated with \citet{2014A&A...571A..16P} cosmology. \citet{2015MNRAS.451.2663H} generated two light cones by stacking the simulation box ($500 h^{-1}$~Mpc per side) with no additional transformations. One light cone places the observer at the origin and the other places the observer at $x=y=z=250 h^{-1}$~Mpc. The light cone galaxy catalogs include observer-frame magnitudes for 9 bandpasses, including SDSS $g$ and $r$, both of which we use in this study. The mass resolution of the Millennium simulation dark matter particles is  $1.43\times10^{9} h^{-1} $M$_{\odot}$, and the stellar mass resolution of the luminous galaxies is $\sim 10^{7}$ M$_{\odot}$. The light cone galaxy catalogs are publicly available via the \texttt{Henriques2015a} table in the Cosmology Science Domain on the SDSS SciServer.

\begin{figure*}
    \centering
   \includegraphics[width=0.73\linewidth]{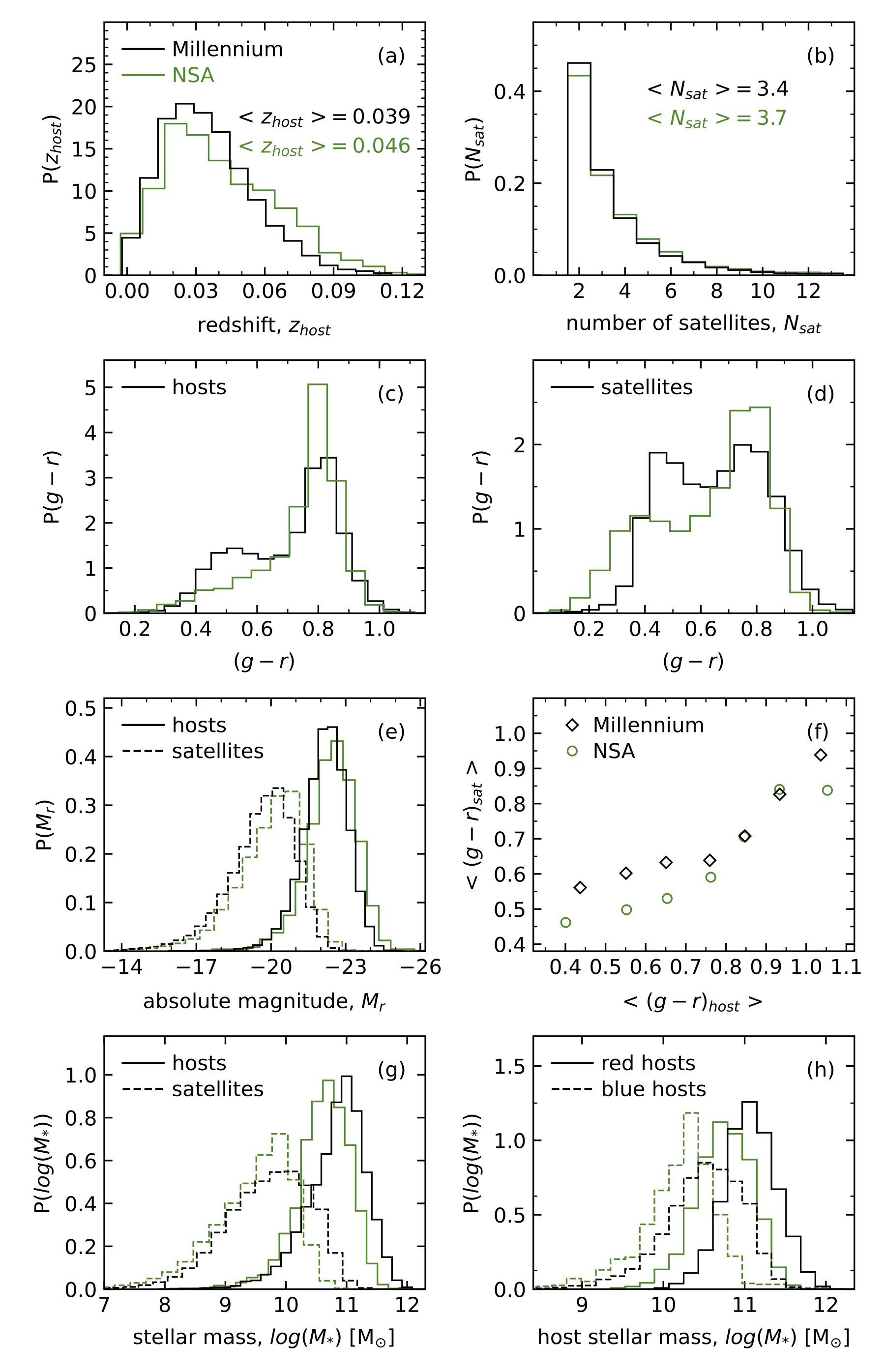}
    \caption{Properties of the Millennium (black) and NSA (green) samples. (a) Host redshift distributions. (b) Number of satellites per system, truncated for clarity. (c) Host rest-frame $(g - r)$ color distributions. (d) Satellite rest-frame $(g - r)$ color distributions. (e) Host and satellite $r$-band absolute magnitude distributions. (f) Dependence of mean satellite color on host color. (g) Host and satellite stellar mass distributions. (h) Host stellar mass distributions, computed separately for red and blue hosts.}
    \label{fig:sample_properties}
\end{figure*}

In order to facilitate direct comparison with the NSA sample from BS20, we restrict our analysis to Millennium galaxies with $r < 17.77$, consistent with the SDSS spectroscopic completeness limit. Furthermore, host and satellite galaxies were selected using the same redshift space criteria as BS20. Potential host galaxies were required to have radial velocities $v > 500$~km~s$^{-1}$ and to be at least 1~mag brighter than all other galaxies within a projected distance $r_{p} < 700$~kpc and radial velocity difference $|dv| < 1000$~km~s$^{-1}$. Additionally, as in BS20, potential host galaxies were required to have apparent magnitudes in the range $9 < r < 15$, which ensures that the majority of hosts (90.2\%) have luminosities brighter than $L_{*}$ (i.e., $M_{r,*} = -20.83$; \citealt{2001AJ....121.2358B}). Satellite galaxies were required to be located within a projected distance $r_{p} < 500$~kpc of their hosts and a radial velocity difference $|dv| < 500$~km~s$^{-1}$. Since we are interested in analysing the pairwise clustering of the satellites, we further restricted the sample to systems for which two or more satellites were identified.  
Below, we refer to the host-satellite sample drawn from the mock all-sky redshift surveys as the ``Millennium sample'' and to the host-satellite sample drawn from BS20 as the ``NSA sample''. Various properties of both samples are summarized in Figure~\ref{fig:sample_properties} and Table~\ref{table:sample_properties}. 

From Table~\ref{table:sample_properties}, the Millennium sample is considerably larger than the NSA sample ($\sim 7$ times more hosts and $\sim 6$ times more satellites).  From Table~\ref{table:sample_properties} and
Figure~\ref{fig:sample_properties}, the majority of the properties are similar in the two samples (i.e., redshift distribution, number of satellites per system, extinction-corrected and $k$-corrected absolute magnitudes of the hosts and satellites, as well as the dependence of the mean satellite color on host color).  Compared to the NSA sample, the Millennium hosts are somewhat more intrinsically luminous than their satellites (i.e., the mean host-to-satellite luminosity ratio is $\sim 40\%$ larger for the Millennium sample; see Table~\ref{table:sample_properties}).  In addition, the Millennium hosts and satellites have somewhat larger stellar masses than those in the NSA sample (i.e., the median stellar mass for hosts and satellites is $\sim 1.4$ times larger in the Millennium sample). Here the stellar masses for the NSA galaxies correspond to the \texttt{elpetro\_mass} parameter in the NSA database. While there is a systematic offset between the stellar mass distributions of the NSA and Millennium hosts, the difference between the stellar masses as a function of host color is consistent between the samples.  That is, as expected, the red hosts are somewhat more massive than the blue hosts and in both samples the median and mean stellar mass of the red hosts exceeds that of the blue hosts by a factor of $\sim 3$ (see Table~\ref{table:sample_properties}).  Lastly, from Figure~\ref{fig:sample_properties} (panels c and d), the two samples differ somewhat in terms of the distribution of $(g-r)$ colors for both the hosts and the satellites. The Millennium hosts show a pronounced color bi-modality that is not shown by the NSA hosts and the fraction of blue satellites is considerably larger in the Millennium sample.

\begin{deluxetable}{ccc}
\tablenum{1}
\tablecaption{Host-Satellite Samples}
\tablewidth{0pt}
\tablehead{property & Millennium & NSA}
\startdata
number of hosts & 23,652 &  3,575 \\
number of satellites & 80,703 & 13,090 \\
mean number of satellites & 3.4 & 3.7 \\
median number of satellites & 3 & 3 \\
maximum number of satellites & 46 & 26 \\
mean host redshift & 0.039 & 0.046 \\
median host redshift & 0.037 & 0.042 \\
median host $M_{r}$ & -22.0 & -22.8 \\
median satellite $M_{r}$ & -19.6 & -19.9 \\
mean host-to-satellite luminosity ratio & 27.3 & 19.9 \\
mean host stellar mass $[M_{\odot}]$ & 1.3x$10^{11}$ & 9.2x$10^{10}$ \\
median host stellar mass $[M_{\odot}]$ & 9.6x$10^{10}$ & 6.7x$10^{10}$ \\
mean satellite stellar mass $[M_{\odot}]$ & 1.7x$10^{10}$ & 9.2x$10^{9}$ \\
median satellite stellar mass $[M_{\odot}]$ & 7.8x$10^{9}$ & 5.3x$10^{9}$ \\
mean red host stellar mass $[M_{\odot}]$ & 1.9x$10^{11}$ & 1.1x$10^{11}$ \\
median red host stellar mass $[M_{\odot}]$ & 1.4x$10^{11}$ & 8.1x$10^{10}$ \\
mean blue host stellar mass $[M_{\odot}]$ & 6.3x$10^{10}$ & 3.6x$10^{10}$ \\
median blue host stellar mass $[M_{\odot}]$ & 4.2x$10^{10}$ & 2.5x$10^{10}$ \\
mean red satellite stellar mass $[M_{\odot}]$ & 3.3x$10^{10}$ & 1.4x$10^{10}$ \\
median red satellite stellar mass $[M_{\odot}]$ & 2.4x$10^{10}$ & 1.1x$10^{10}$ \\
mean blue satellite stellar mass $[M_{\odot}]$ & 6.4x$10^{9}$ & 3.3x$10^{9}$ \\
median blue satellite stellar mass $[M_{\odot}]$ & 3.2x$10^{9}$ & 1.6x$10^{9}$ \\
number of red hosts & 13,385 & 2,719 \\
number of blue hosts & 10,267 & 856 \\
number of red satellites & 33,125 & 6,873 \\
number of blue satellites & 47,578 & 6,217 \\
\enddata
\end{deluxetable}
\label{table:sample_properties}

In order to subdivide the Millennium sample into ``red'' and ``blue'' galaxies, we fit the distribution of the intrinsic $(g - r)$ colors of all $z = 0$ Millennium galaxies (i.e., not just the galaxies that were selected as hosts and satellites) with the sum of two Gaussian distributions. From this, we find that the Gaussians intersect at $(g - r) = 0.74$ and we therefore take Millennium galaxies with $(g - r) < 0.74$ to be ``blue'' and those with $(g - r) \geq 0.74$ to be ``red''. 
In the case of the NSA sample, the division between red and blue galaxies was taken to be $(g - r) = 0.7$ (e.g., \citealt{2003ApJ...594..186B}).  Using these color divisions, we find that, overall, the Millennium sample is bluer than the NSA sample. From Table~\ref{table:sample_properties}, the fraction of Millennium hosts that are blue in color (43.4\%) is considerably higher than the fraction of NSA hosts that are blue in color (23.9\%).   In addition, the fraction of satellites that are blue in color is somewhat higher for the Millennium sample (59.0\%) than it is for the NSA sample (50.9\%).

\section{Methods and Results} \label{sec:Results}

We use two complementary metrics to quantify the degree of lopsidedness in the spatial distributions of the satellites. First, we use the same metric as BS20: the probability distribution of the polar angle differences between pairs of satellites, $P(\Delta \phi)$, aggregated over all systems. Second, we use the Mean Resultant Length, which quantifies the directionality of a distribution of points (i.e., the directionality, relative to the host, of all satellites in a given system).

\subsection{Pairwise Polar Angle Differences}

As in BS20, we construct probability distributions for the polar angle differences between each pair of satellites in a system, $P(\Delta \phi)$.  Here $\Delta\phi$ is defined such that pairs of satellites with $\Delta\phi \sim 0^{\circ}$ are located on the same side of their host, and pairs of satellites with $\Delta\phi \sim 180^{\circ}$ are located on opposite sides of their host (see Figure~3 of BS20 for an illustration of the definition of $\Delta\phi$). To make a fair comparison between the Millennium and NSA samples, we also restrict the maximum number of satellites in a given system to be the lowest maximum number of satellites between the two samples.  That is, the maximum number of satellites around blue hosts in the Millennium sample is 22, while in the NSA sample the maximum number of satellites around blue hosts is 17.  So, when comparing results for the satellites of blue hosts, we restrict both samples to systems that contain at most 17 satellites. Furthermore, to ensure that the results are not affected by different numbers of systems containing $n$ satellites, we compare the NSA sample to 1,000 bootstrap resamplings of the Millennium sample, each containing the same number of systems with $n$ satellites as the NSA sample (i.e., each Millennium sample has the same number of systems with 2 satellites, 3 satellites, 4 satellites, etc. as the NSA sample).

\begin{figure*}
    \centering
    \includegraphics[width=0.95\linewidth]{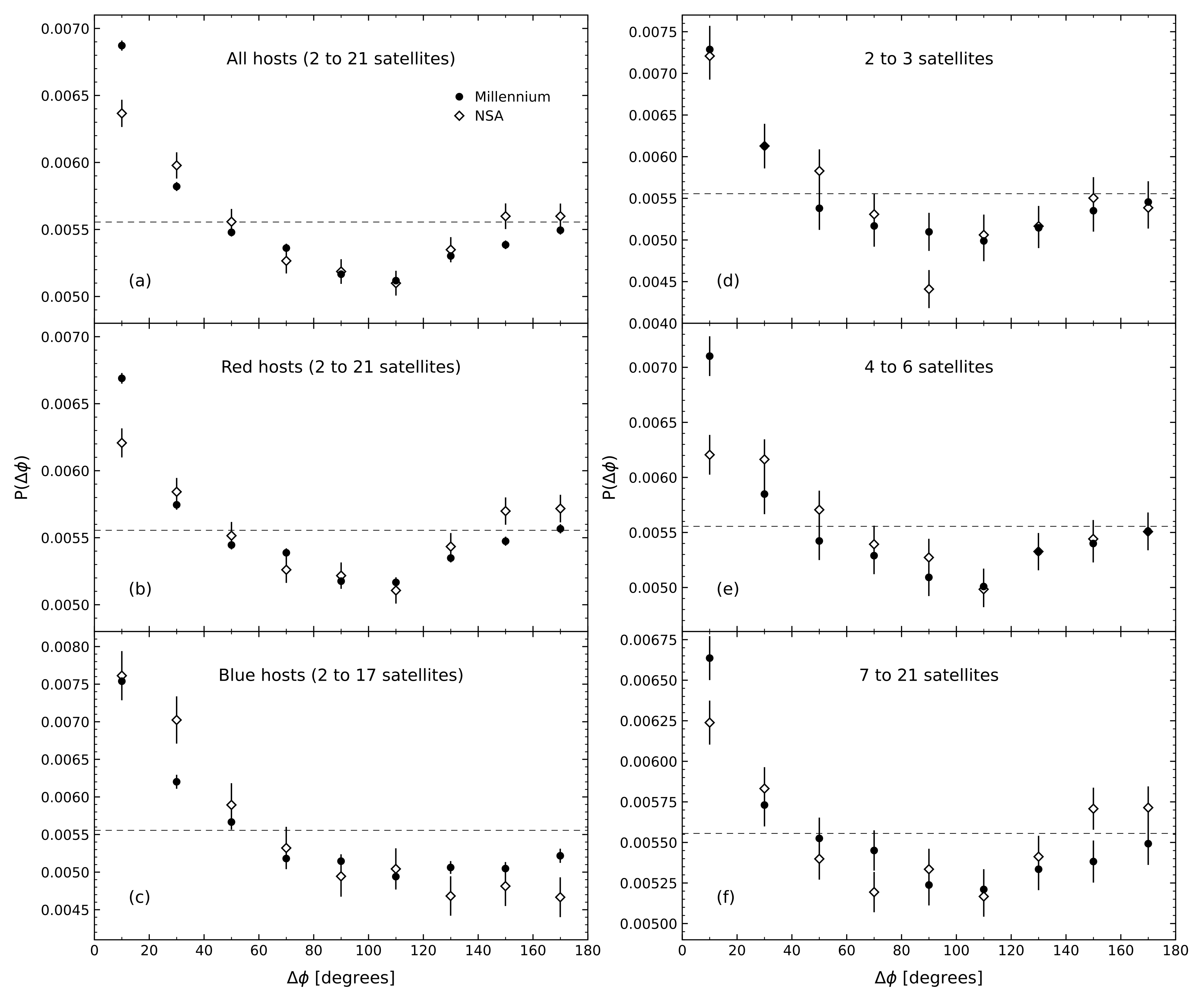}
    \caption{Probability distributions, $P(\Delta \phi)$, for the polar angle difference between pairs of Millennium satellites (circles) and NSA satellites (diamonds). Note that the range of ordinate values differs amongst the panels. (a) All host-satellite systems. (b) Satellites of red hosts. (c) Satellites of blue hosts. (d) Systems with 2 to 3 satellites. (e) Systems with 4 to 6 satellites. (f) Systems with 7 to 21 satellites. 
    Dashed lines show $P(\Delta \phi)$ for a uniform circular distribution of satellites.}
    \label{fig:p_of_delta_phi}
\end{figure*}

Figure \ref{fig:p_of_delta_phi} shows $P(\Delta \phi)$ for various subsets of the Millennium and NSA samples. Dashed lines show $P(\Delta \phi)$ for a uniform circular distribution of satellites. Here, $1\sigma$ error bars were generated from 10,000 bootstrap resamplings of the ensemble set of $\Delta \phi$ values for each sample, and error bars are omitted from figures when they are smaller than the sizes of the data points.

From Figure~\ref{fig:p_of_delta_phi}, pairs of satellites in the Millennium sample show a strong tendency to be located on the same side of their hosts, resulting in lopsided distributions.  The result is highly significant with $\chi^2$ tests indicating that, in all six panels of Figure~\ref{fig:p_of_delta_phi},
$P(\Delta\phi)$ for the Millennium sample is inconsistent with a uniform circular distribution at confidence levels $> 99.9999\%$.  In addition, in panels a), b), e), and f) (i.e., complete sample, satellites of red hosts, systems with 4 to 6 satellites, and systems with 7 to 21 satellites), $\chi^2$ tests indicate that the distribution of Millennium satellites is inconsistent with the distribution of NSA satellites at confidence levels $> 99.5\%$.  The sense of the disagreement between the two samples is that the Millennium sample shows a greater degree of lopsidedness than the NSA sample.  In the case of systems with blue hosts and systems with 2 to 3 satellites (which are not fully independent, since blue hosts tend to have fewer satellites than red hosts), the lopsidedness of the satellite distribution is the greatest, and $\chi^2$ tests indicate that the Millennium and NSA results for $P(\Delta\phi)$ are consistent with one another. 

\begin{figure}
    \centering
    \includegraphics[width=\columnwidth]{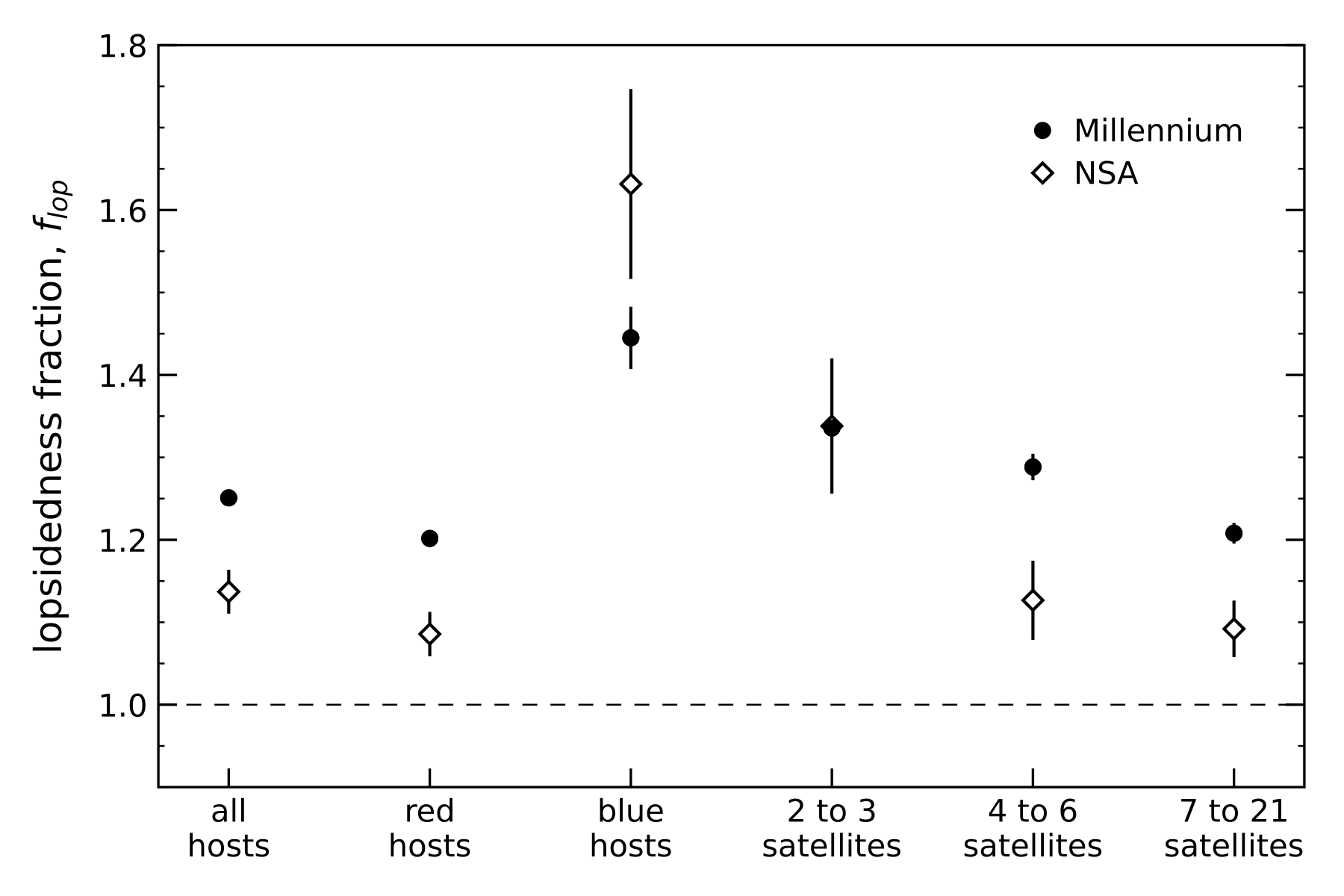}
    \caption{Lopsidedness fractions, $f_{\rm lop}$, for the host-satellite samples.  Circles: Millennium sample. Diamonds: NSA sample.
    }
    \label{fig:f_lop}
\end{figure}

We further characterize the lopsidedness of the satellite distributions by computing the ratio of the number of pairs of satellites with $0^\circ \le \Delta\phi \le 20^\circ$ and $160^\circ \le \Delta\phi \le 180^\circ$ (i.e., the ratio of the number of pairs of satellites on the same side of their host to the number of pairs on opposite sides of their host).  We refer to this ratio as the lopsidedness fraction, $f_{\rm lop}$, the results of which are shown in Figure~\ref{fig:f_lop}.  From Figure~\ref{fig:f_lop}, the lopsidedness fraction for the Millennium satellites significantly exceeds the lopsideness fraction for the NSA satellites in the complete sample, in systems with red hosts, in systems with 4 to 6 satellites, and in systems with 7 to 21 satellites.  For systems with only 2 to 3 satellites, the lopsideness fraction is identical in both samples.  For systems with blue hosts, the lopsideness fraction for the Millennium satellites is somewhat lower than it is for the NSA satellites.

\subsection{Mean Resultant Length (MRL)}

The MRL is a circular summary statistic that measures the directionality of a distribution of points.  For a host-satellite system with $n$ satellites, and satellite polar angles measured in a coordinate system centered on the host, the MRL is defined as
\begin{equation}
\overline{R}=\frac{1}{n} \sqrt{
\left( \sum_{i=1}^{n} \cos \phi_{i} \right)^2 +
\left( \sum_{i=1}^{n} \sin \phi_{i} \right)^2
}
\end{equation}
where $\phi_i$ are the polar angles of the satellites (see, e.g., \citealt{2012msma.book.....F}). 
If the satellites are clustered towards one direction relative to the host, $\overline{R} \rightarrow 1$.  If the satellites are distributed in a balanced way about the host, $\overline{R} \rightarrow 0$. The MRL can be interpreted as the magnitude of the vector sum of individual vectors, each with length $1/n$ and each pointing in the direction of one of the satellites, whose length would be unity if all $n$ satellites had the same polar angle.  

A key feature of the MRL is that it takes into account the locations of {\it all} satellites in a given system, not merely {\it pairs} of satellites in the system (i.e., as is the case with the probability distribution of polar angle differences in the previous subsection).  Furthermore, the value of $\overline{R}$ within a system gives an indication of whether the satellites in \textit{that particular system} are distributed in a lopsided or balanced way, as opposed to $P(\Delta \phi)$, which only measures the aggregate distribution of polar angle differences between pairs of satellites within the sample (i.e., not the degree to which individual systems in the sample show lopsided satellite distributions).

\begin{figure}
    \centering
    \includegraphics[width=\columnwidth]{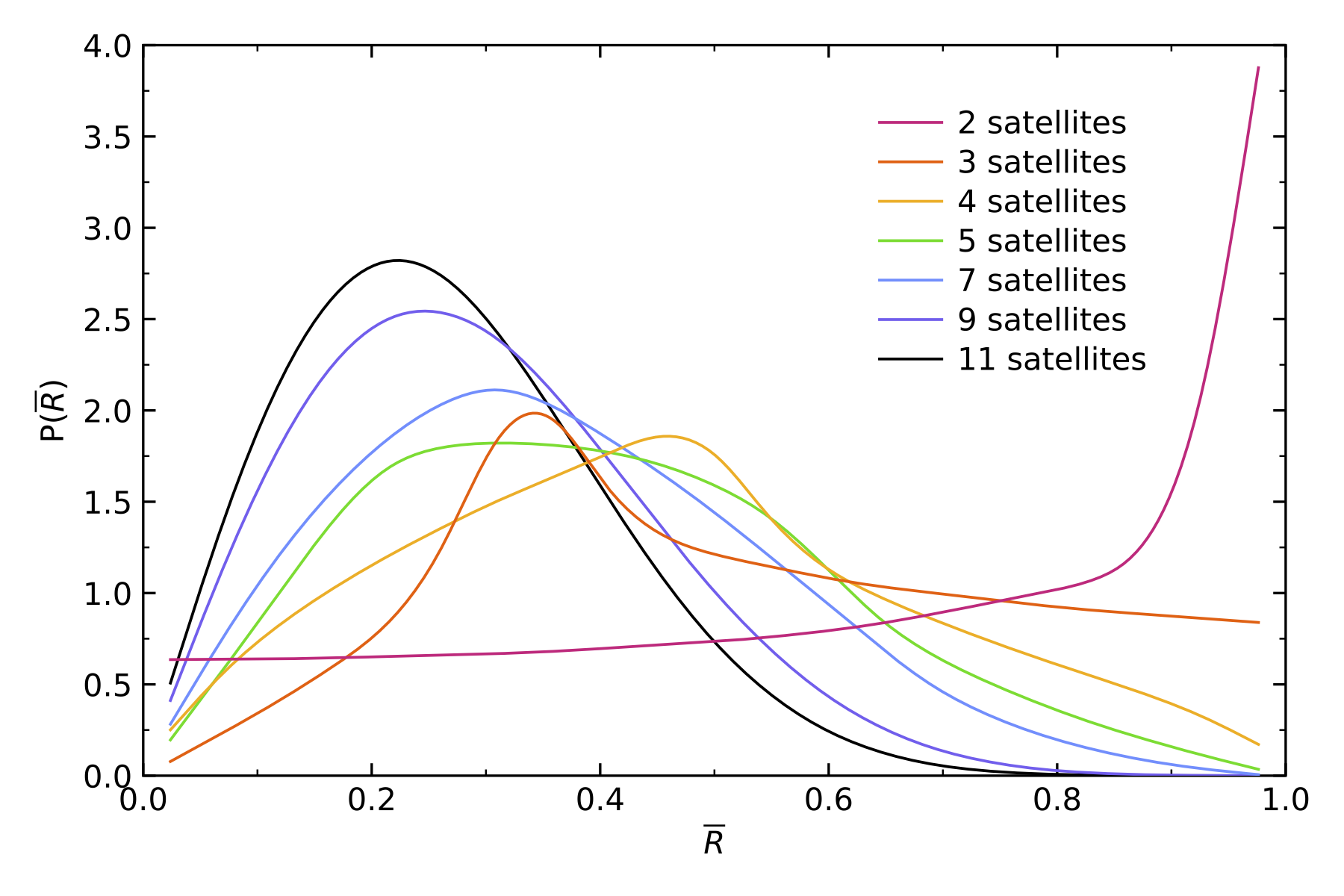}
    \caption{Expected probability distributions, P$(\overline{R})$, for the Mean Resultant Length, $\overline{R}$, in systems that contain $n$ randomly distributed satellite galaxies.}
    \label{fig:mrl_random}
\end{figure}

Figure \ref{fig:mrl_random} shows the expected probability distributions, $P(\overline{R})$, for randomly distributed satellites in systems with various numbers of satellites.  From this, it is clear that the shape of $P(\overline{R})$ depends strongly on the number of satellites per system. Therefore, throughout our analysis below, we are careful to compare samples that contain an identical number of systems with $n$ satellites.  To compare results for the NSA sample to results from the Millennium sample, we again use 1,000 bootstrap resamplings of the Millennium sample, each of which contains the same number of systems with $n$ satellites as the NSA sample.  To compare results for the Millennium and NSA samples to the expected distribution for randomly distributed satellites, we also match the number of systems with $n$ randomly distributed satellites to the number of systems with $n$ satellites in the NSA sample.

\begin{figure*}
    \centering
    \includegraphics[width=0.95\linewidth]{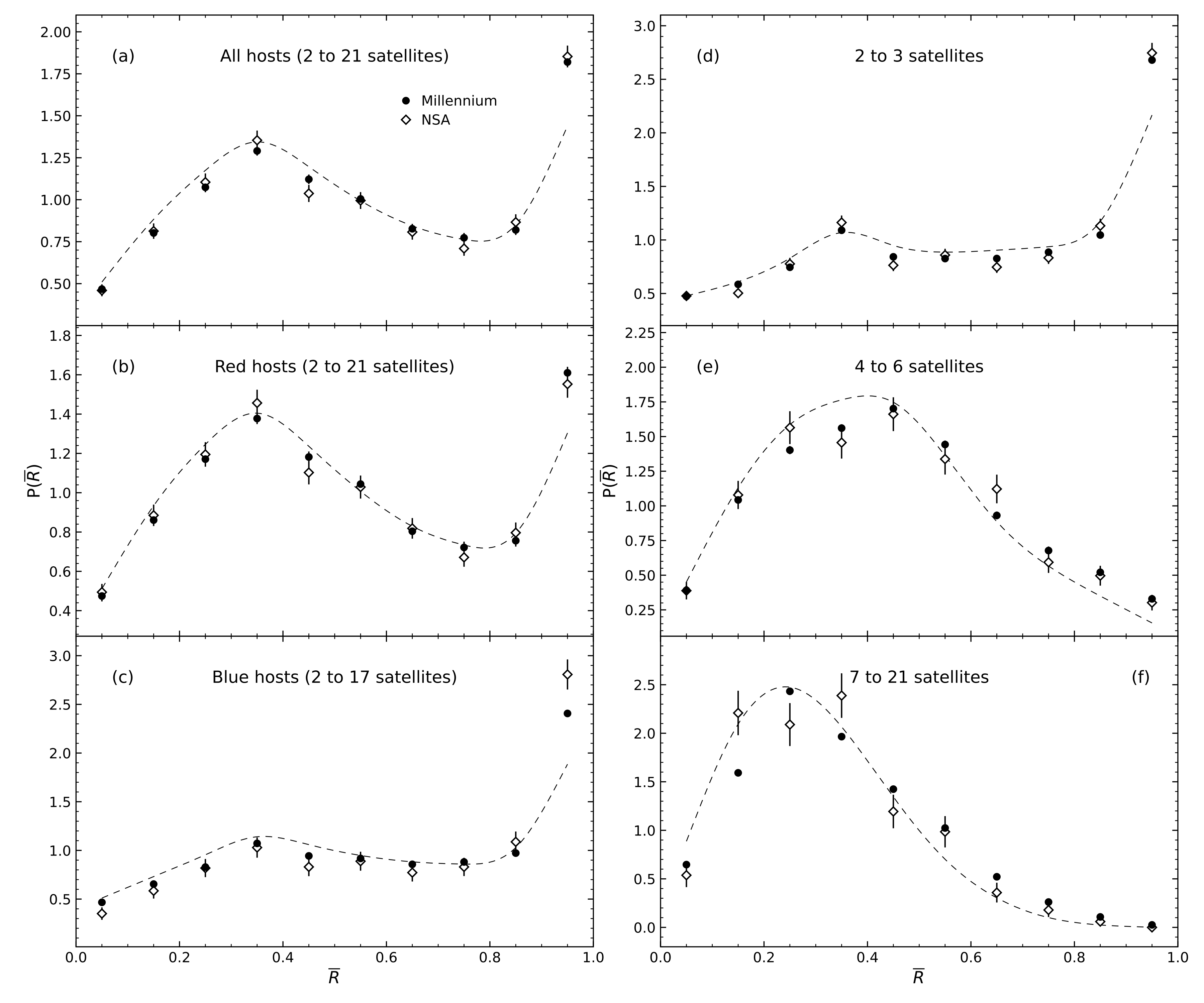}
    \caption{Probability distributions, P$(\overline{R})$, for the Mean Resultant Length of each host-satellite system.  Note that the range of ordinate values differs amongst the panels. Circles: Millennium satellites.  Diamonds: NSA satellites. Dashed lines: expected distributions for samples with random satellite locations. (a) All host-satellite systems. (b) Satellites of red hosts. (c) Satellites of blue hosts. (d) Systems with 2 to 3 satellites. (e) Systems with 4 to 6 satellites. (f) Systems with 7 to 21 satellites.}
    \label{fig:p_of_rbar}
\end{figure*}

Figure~\ref{fig:p_of_rbar} shows $P(\overline{R})$ for various subsets of the Millennium and NSA samples (circles and diamonds, respectively) as well as results for 10,000 iterations of the sample with randomized values of $\phi$ (dashed line). For clarity of the plot (i.e., to highlight the differences between the expectations for randomly distributed satellites vs.\ the Millennium and NSA satellites), Figure~\ref{fig:p_of_rbar} displays the results for $P(\overline{R})$ for the randomly distributed satellites as spline fit lines through the values at the centers of each of the bins. Here, $1\sigma$ error bars were generated from 10,000 bootstrap resamplings of the ensemble set of $\overline{R}$ values for each sample.

It is clear from Figure~\ref{fig:p_of_rbar}a) that the distributions of $\overline{R}$ for the full Millennium and NSA samples show similar levels of directionality, as well as significantly greater directionality than the sample with randomized satellite locations (i.e., more systems with $\overline{R}$ values close to 1). Similar to our results for the dependence of $P(\Delta\phi)$ on host color in Figure~\ref{fig:p_of_delta_phi}, Figure~\ref{fig:p_of_rbar} indicates that the satellites of blue hosts (panel c) show greater directionality in their locations than do the satellites of red hosts (panel b).  Further, from Figure~\ref{fig:p_of_rbar}d), the lopsidedness in both the Millennium sample and the NSA sample is driven largely by systems with 2 to 3 satellites.  When we compare the results for the Millennium satellites and the NSA satellites in Figure~\ref{fig:p_of_rbar}, $\chi^2$ tests indicate that, with the exception of systems with 7 to 21 satellites, $P(\overline{R})$ for both samples is consistent with being drawn from the same underlying distribution (i.e., as quantified by the probability distribution of MRL values, the samples show the same degree of lopsidedness).  In the case of systems with 7 to 21 satellites, a $\chi^2$ test shows that the probability distributions for the two samples are moderately inconsistent with being drawn from the same underlying distribution (confidence level of 99.72\%, corresponding to a reduced $\chi^2$ value of 2.80).

\subsection{Interlopers}

In an observational sample of host galaxies and their satellites, line of sight distances are rarely known.  Because of this, host-satellite systems are obtained using redshift space selection criteria that are based on proximity in the plane of the sky (i.e., angular separation) and proximity in line-of-sight velocity.  The majority of the objects that are identified as satellite galaxies are, in fact, legitimate satellites.  However, some fraction of the objects that are identified as satellites are actually objects that pass the selection criteria but are not physically associated with the central, host galaxy.  Such objects are known as ``interlopers'' (or ``false satellites'').  In an observational sample it is, of course, not possible to distinguish legitimate satellites from interlopers.  In a sample obtained from a simulation it is, however, possible to determine which objects are interlopers since, in addition to the mock redshift surveys, full 3D spatial information is known for each galaxy.

Here we investigate the effects of interlopers on our detections of lopsidedness in the Millennium satellite distribution, with an emphasis on determining whether the lopsidedness is solely attributable to the presence of interlopers.  We define legitimate satellites to be those Millennium objects that pass the redshift space selection criteria for satellites and are also located within the 3D virial radius of their host galaxy. From this, we find that 60.8\% of the Millennium satellites are legitimate satellites. The fraction of objects that qualify as legitimate satellites in a simulation will vary with both the redshift space selection criteria and the maximum host-satellite distance in 3D, but the interloper fraction is typically of order one-third of the objects.  The results for our sample are reasonably consistent with the results from \citet{2010ApJ...709.1321A}, who used an earlier mock redshift survey of the Millennium simulation to obtain a sample of isolated host galaxies and their satellites.  \citet{2010ApJ...709.1321A} used somewhat different selection criteria and defined legitimate satellites to be those objects that were located within a 3D distance of 500 kpc from their host. From this, \citet{2010ApJ...709.1321A} found a legitimate satellite fraction of 66.8\%. 

\begin{figure*}
    \centering
    \includegraphics[width=0.95\linewidth]{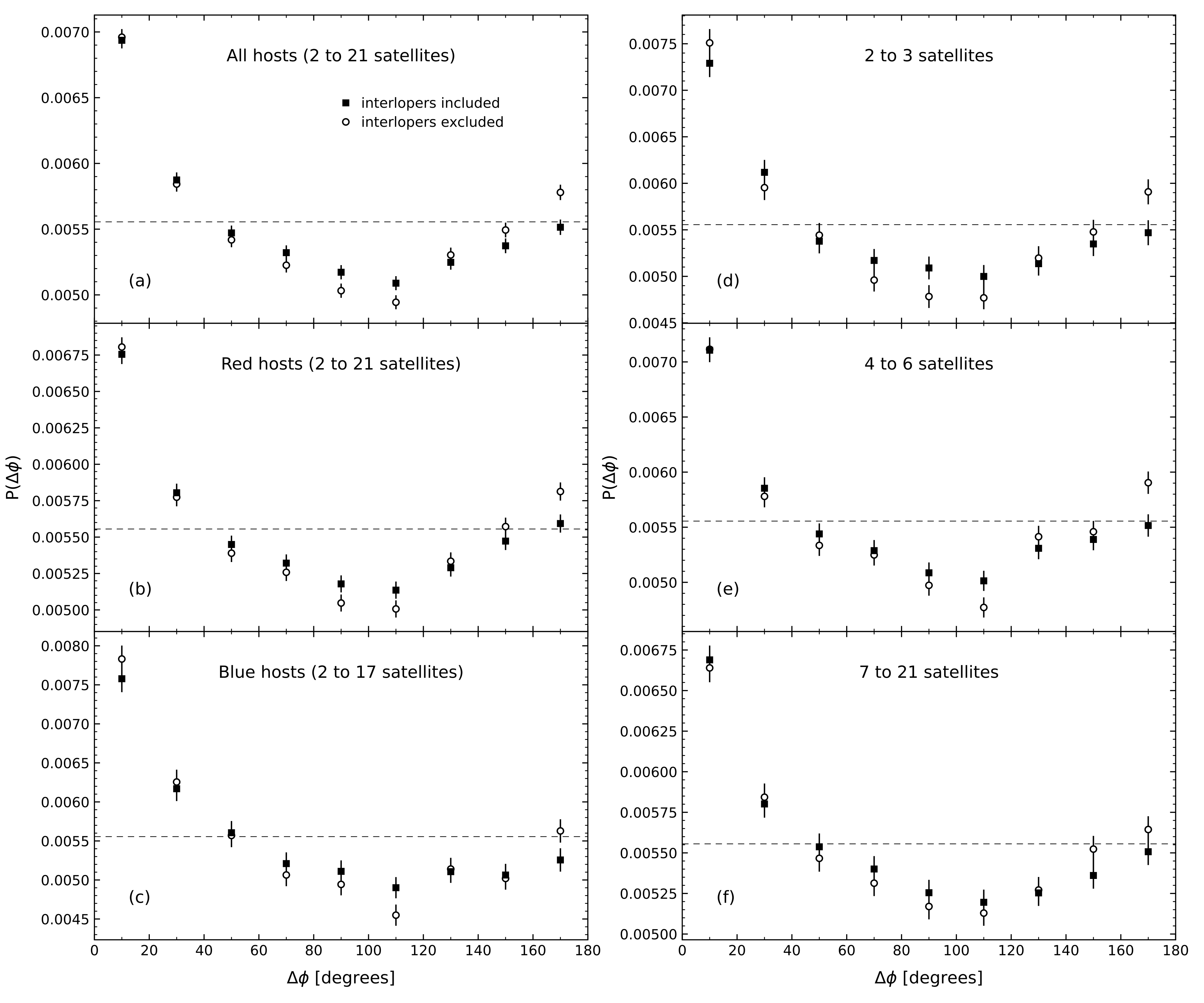}
    \caption{Probability distributions, $P(\Delta \phi)$, for the Millennium sample.  Circles: legitimate satellites only (see text for definition). Squares: complete satellite samples, including interlopers. (a) All host-satellite systems. (b) Satellites of red hosts. (c) Satellites of blue hosts. (d) Systems with 2 to 3 satellites. (e) Systems with 4 to 6 satellites. (f) Systems with 7 to 21 satellites. Dashed lines show $P(\Delta \phi)$ for a uniform circular distribution.}
    \label{fig:p_of_delta_phi_int}
\end{figure*}

\begin{figure*}
    \centering
    \includegraphics[width=0.95\linewidth]{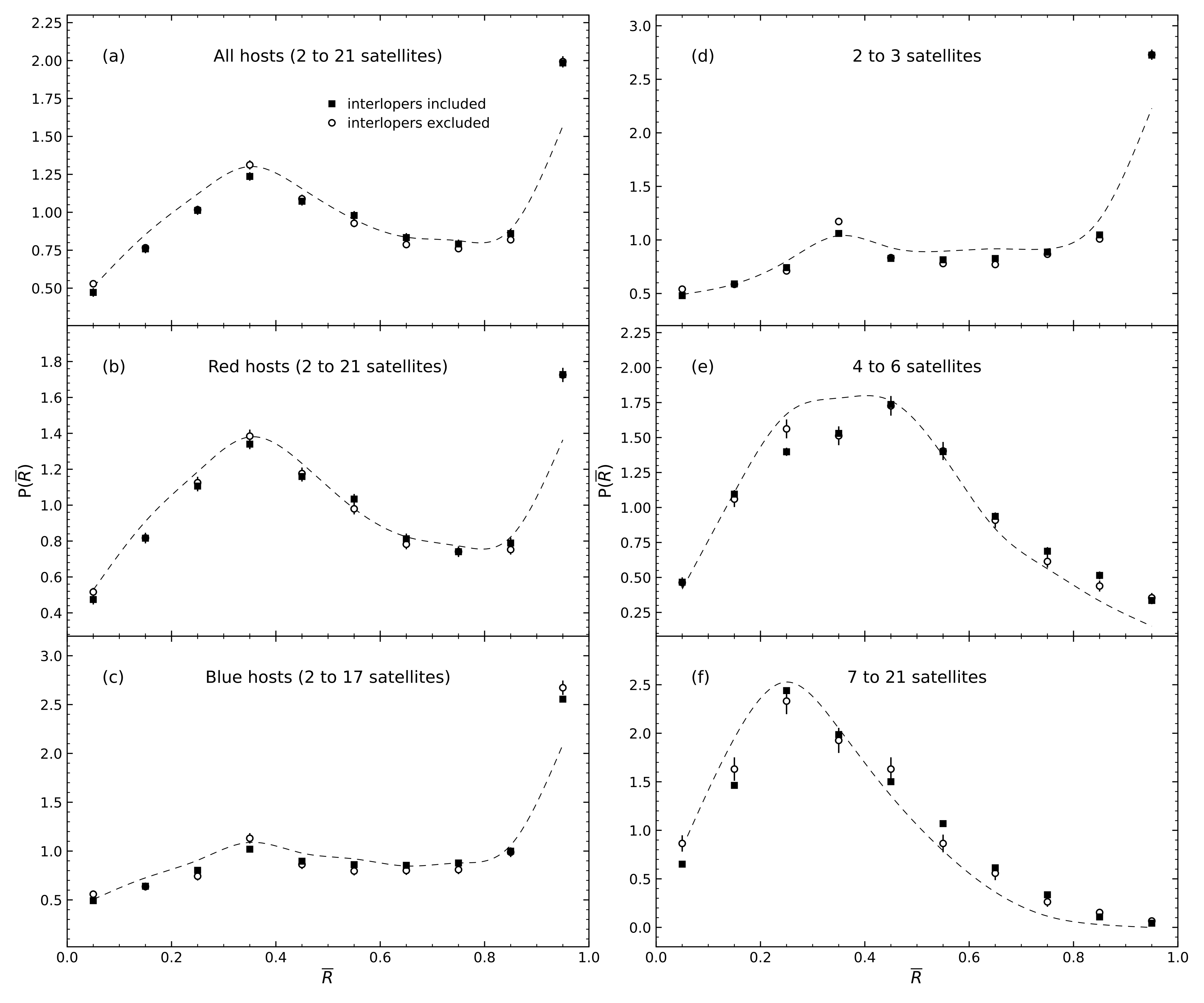}
    \caption{Probability distributions, $P(\overline{R})$, for the Millennium sample.
    Circles: legitimate satellites only (see text for definition).  Squares: complete satellite samples, including interlopers.  Dashed lines: expected distributions for samples with random satellite locations. (a) All host-satellite systems. (b) Satellites of red hosts. (c) Satellites of blue hosts. (d) Systems with 2 to 3 satellites. (e) Systems with 4 to 6 satellites. (f) Systems with 7 to 21 satellites.}
    \label{fig:p_of_rbar_int}
\end{figure*}

Figures~\ref{fig:p_of_delta_phi_int} and \ref{fig:p_of_rbar_int} show the probability distributions, $P(\Delta \phi)$ and $P(\overline{R})$, for legitimate Millennium satellites (``interlopers excluded"; circles) and all Millennium satellites (``interlopers included"; squares). The ``interlopers included'' sample was constructed from 1,000 bootstrap resamplings of the data, each of which contain the same number of systems with $n$ satellites as the ``interlopers excluded'' sample. 

From Figures~\ref{fig:p_of_delta_phi_int} and \ref{fig:p_of_rbar_int}, it is clear that the lopsidedness of the Millennium satellite distribution persists even when interlopers are excluded from the sample.  Therefore, the lopsidedness of the complete Millennium satellite distribution (i.e., with interlopers included) is not solely attributable to the presence of interlopers.  In all cases shown in Figures~\ref{fig:p_of_delta_phi_int} and \ref{fig:p_of_rbar_int}, $\chi^2$ tests indicate that both $P(\Delta \phi)$ and $P(\overline{R})$ for the legitimate Millennium satellites are inconsistent with random satellite distributions at confidence levels $> 99.999\%$.

\begin{figure}
    \centering
    \includegraphics[width=\columnwidth]
    {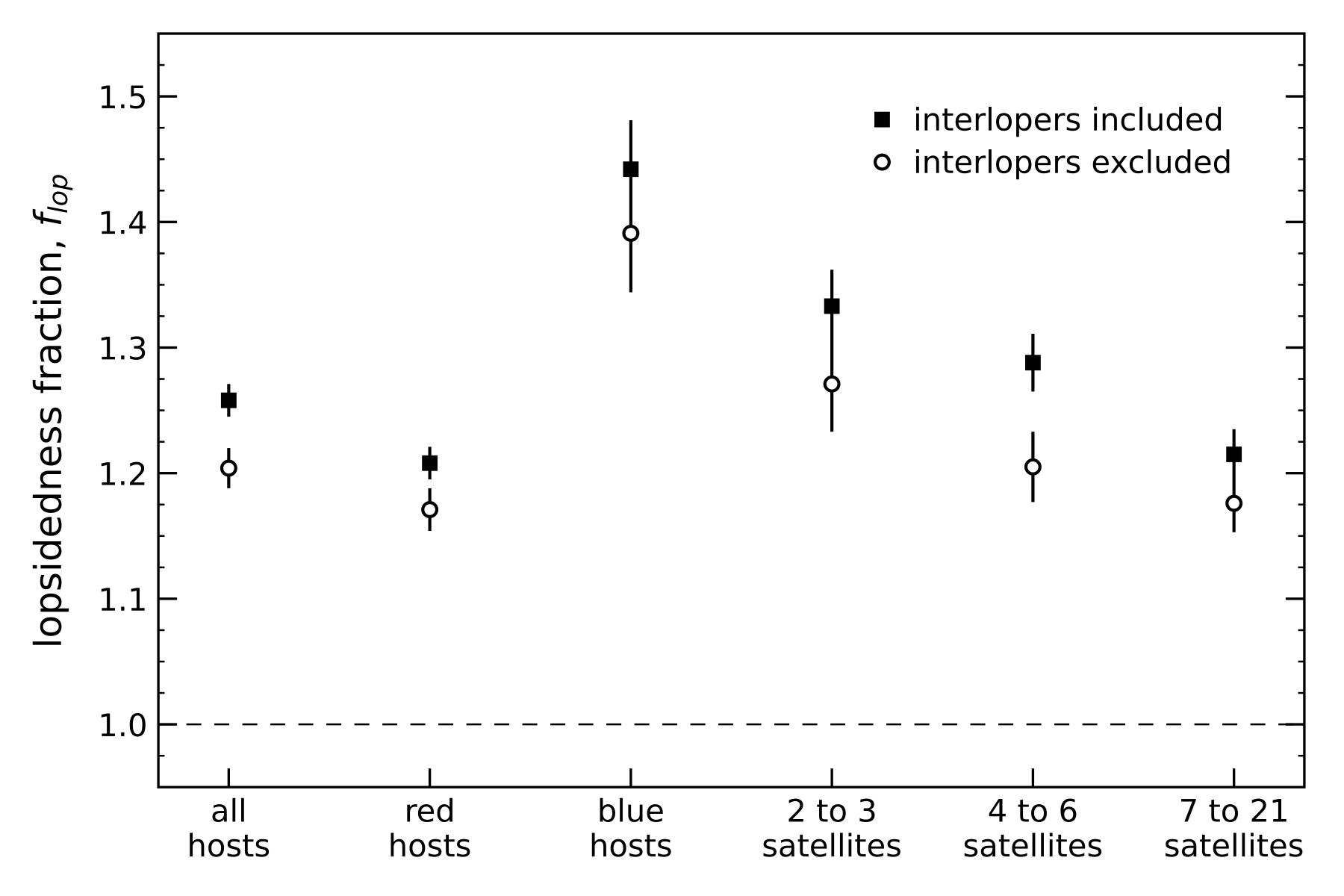}
    \caption{Lopsidedness fractions, $f_{\rm lop}$, for the Millennium sample. Squares: complete satellite samples, including interlopers. Circles: legitimate satellites only.
    }
    \label{fig:f_lop_int}
\end{figure}

Figure~\ref{fig:f_lop_int} shows the lopsidedness fractions for legitimate Millennium satellites and for the the full sample (i.e., with interlopers included).  From this figure, it is clear that the presence of interlopers results in a slight enhancement of the lopsidedness fraction compared to the samples in which interlopers have been excluded.

\begin{figure}
    \centering
    \includegraphics[width=\columnwidth]
    {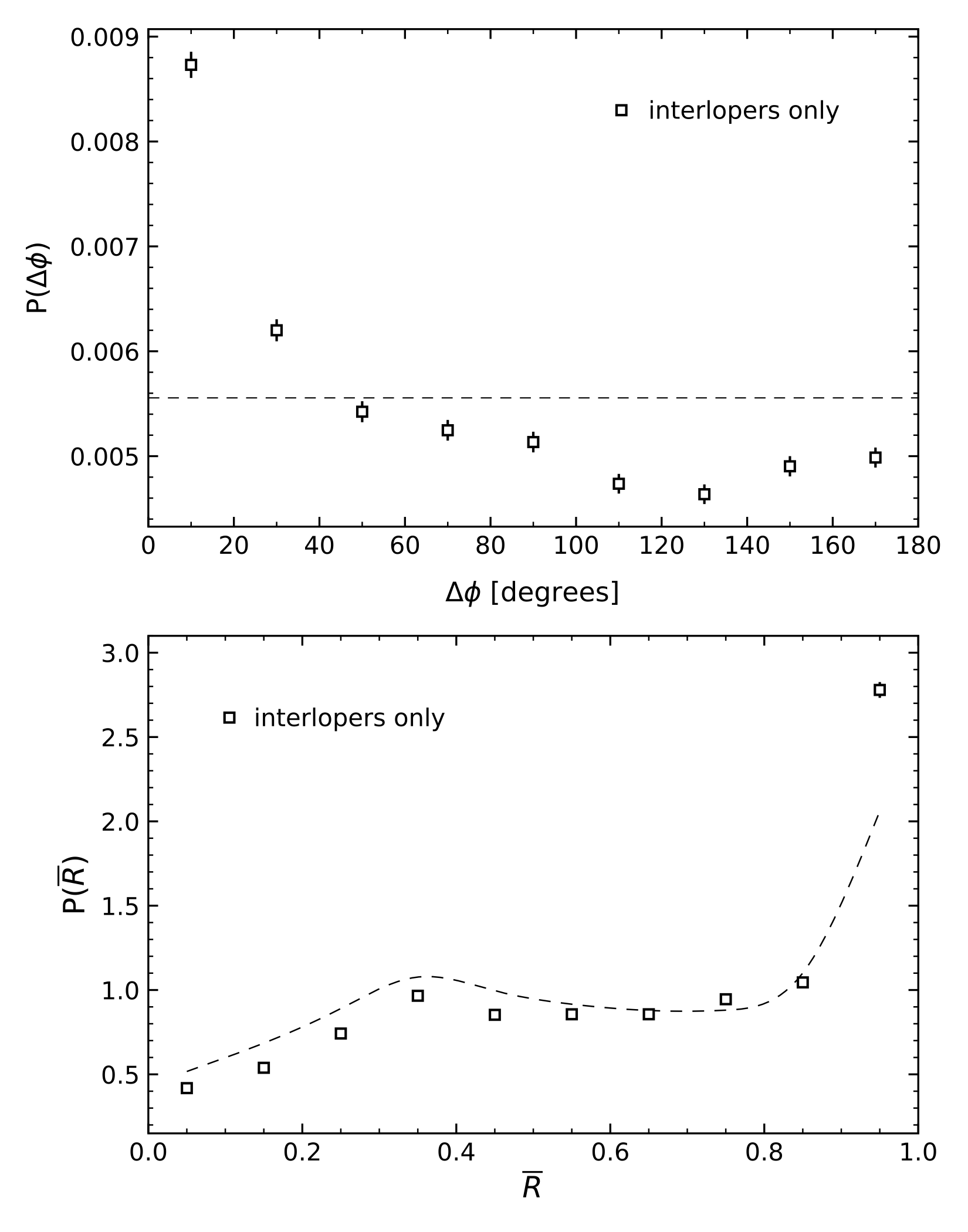}
    \caption{Probability distributions, $P(\Delta \phi)$ and $P(\overline{R})$, computed using only the interlopers that are found in the complete Millennium sample. {\it Top:} $P(\Delta \phi)$, where the dashed line shows the expectation for a uniform circular distribution.  {\it Bottom:} $P(\overline{R})$, where the dashed line shows the expected distribution for a sample with random satellite locations.
    }
    \label{fig:lop_int_only}
\end{figure}

We explore the locations of the interlopers, relative to the hosts, in Figure~\ref{fig:lop_int_only}. The top panel of the figure shows $P(\Delta \phi)$ and the bottom panel shows $P(\overline{R})$.  In both panels, the values of $\Delta \phi$ and $\overline{R}$ were computed solely from the interlopers that were identified using the full sample of Millennium hosts. From Figure~\ref{fig:lop_int_only}, the interlopers show a significant level of  lopsidedness (as measured by both metrics), indicating that interlopers are not a random background. We find that, in 3D, 63\% of the interlopers are located between a host's virial radius and a distance of 500~kpc; that is, the majority of interlopers are not distant galaxies that are completely unrelated to the hosts. Although they adopted a somewhat different definition of ``interloper'' (i.e., objects that were $> 500$~kpc from the host), \citet{2010ApJ...709.1321A} obtained a similar result, finding that the median host-interloper distance was only 630~kpc.  The high level of lopsidedness in the Millennium interloper population suggests that the majority of interlopers are associated with the local large-scale structure that surrounds the hosts and are likely approaching the hosts' halos for the first time, or are on an early fly by.

\subsection{Satellite Infall Redshift}

\begin{figure}
    \centering
    \includegraphics[width=\columnwidth]{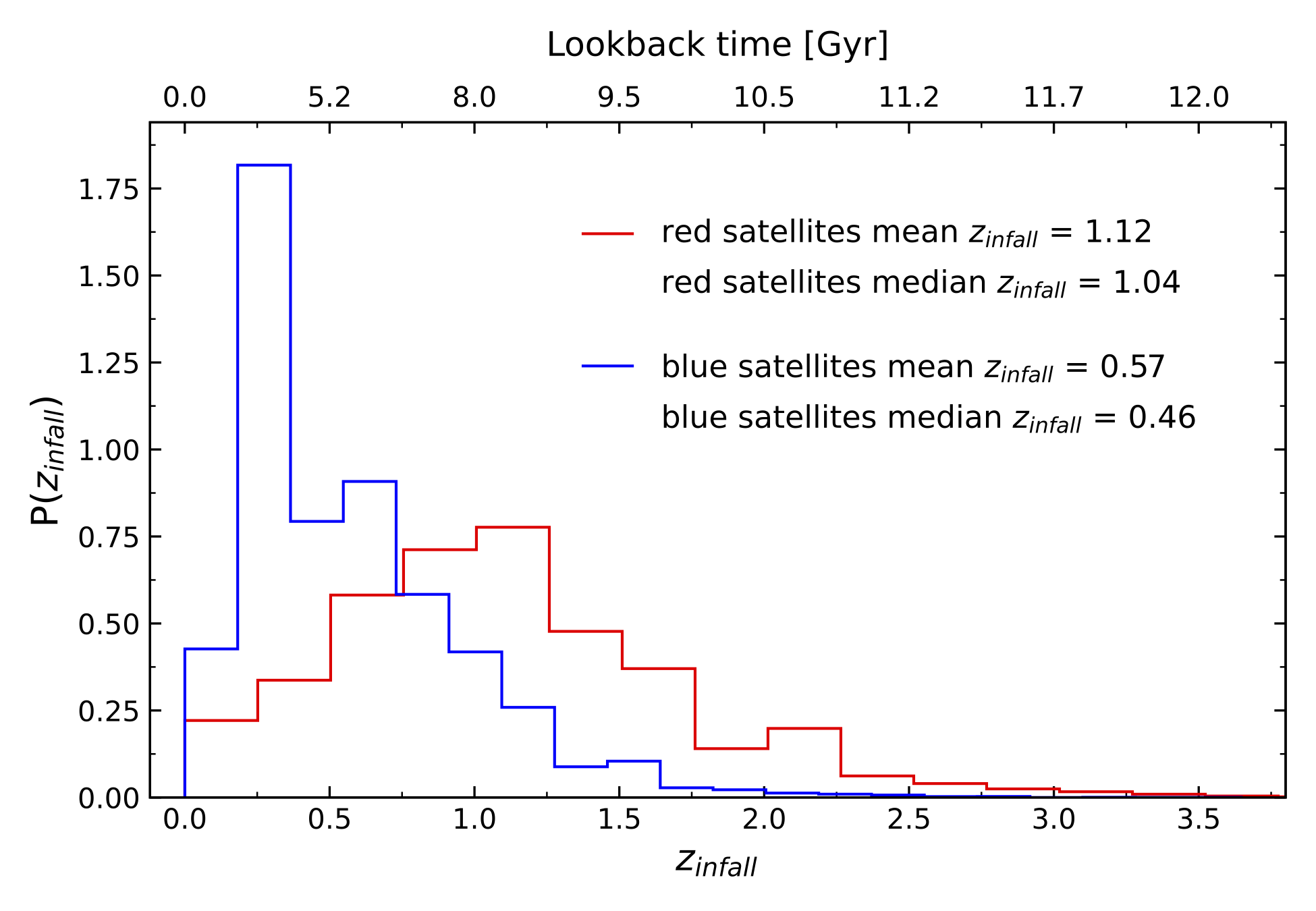}
    \caption{Probability distributions, $P(z_{\rm infall})$, for the infall redshifts of legitimate red and blue Millennium satellites, truncated at high $z$ for clarity.}
    \label{fig:z_infall}
\end{figure}

A potential source of lopsidedness in the satellite distributions is the late-time accretion of small groups of satellites along cosmic filaments. These small groups of 2 to 3 satellites may still be on their first orbit around the host and are unlikely to be a virialized population. To investigate this, we use the infall redshift, $z_{\rm infall}$, at which each legitimate Millennium satellite first entered its host's dark matter halo. Here, $z_{\rm infall}$ is the redshift at which the satellite's distinct halo became a subhalo of the host's halo, as defined by the friend-of-friends algorithm. Figure \ref{fig:z_infall} shows the probability distribution, $P(z_{\rm infall})$, for the infall redshifts of legitimate satellites that are classified as being ``red'' or ``blue'' at the present day. The mean and median infall redshifts of the satellites are listed in Figure \ref{fig:z_infall}.   As expected, most satellites that are red at the present day entered their host's halo earlier than most satellites that are blue at the present day (i.e., there is a difference of approximately 3.2~Gyr in lookback time between the median infall redshifts of the red and blue satellites).

\begin{figure}
    \centering
    \includegraphics[width=\columnwidth]
    {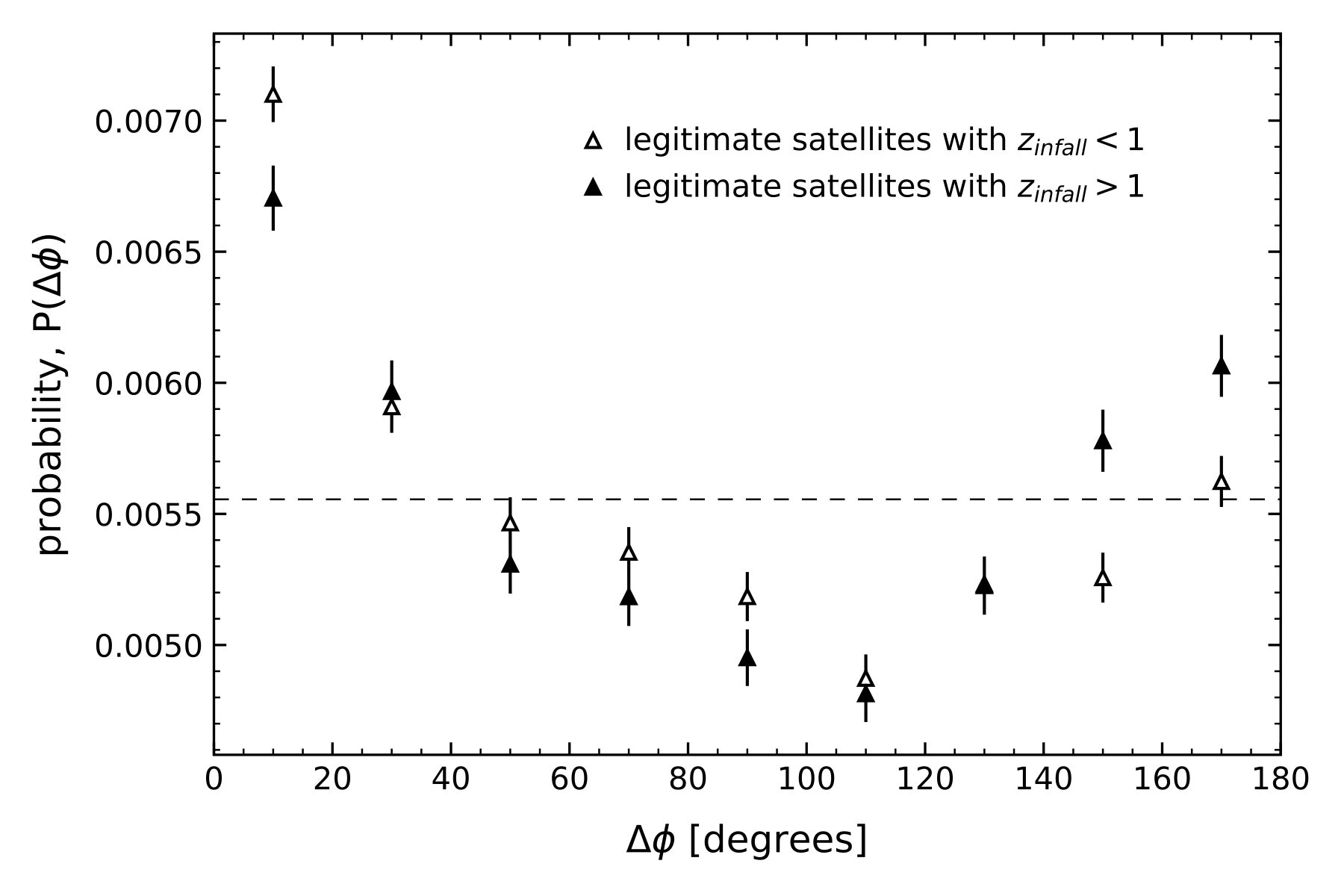}
    \caption{Probability distributions, $P(\Delta \phi)$, for legitimate Millennium satellites. Open triangles: $z_{\rm infall} < 1$.  Filled triangles: $z_{\rm infall} > 1$. Dashed lines: $P(\Delta \phi)$ for a uniform circular distribution.}
    \label{fig:p_of_delta_phi_infall}
\end{figure}

\begin{figure}
    \centering
    \includegraphics[width=\columnwidth]{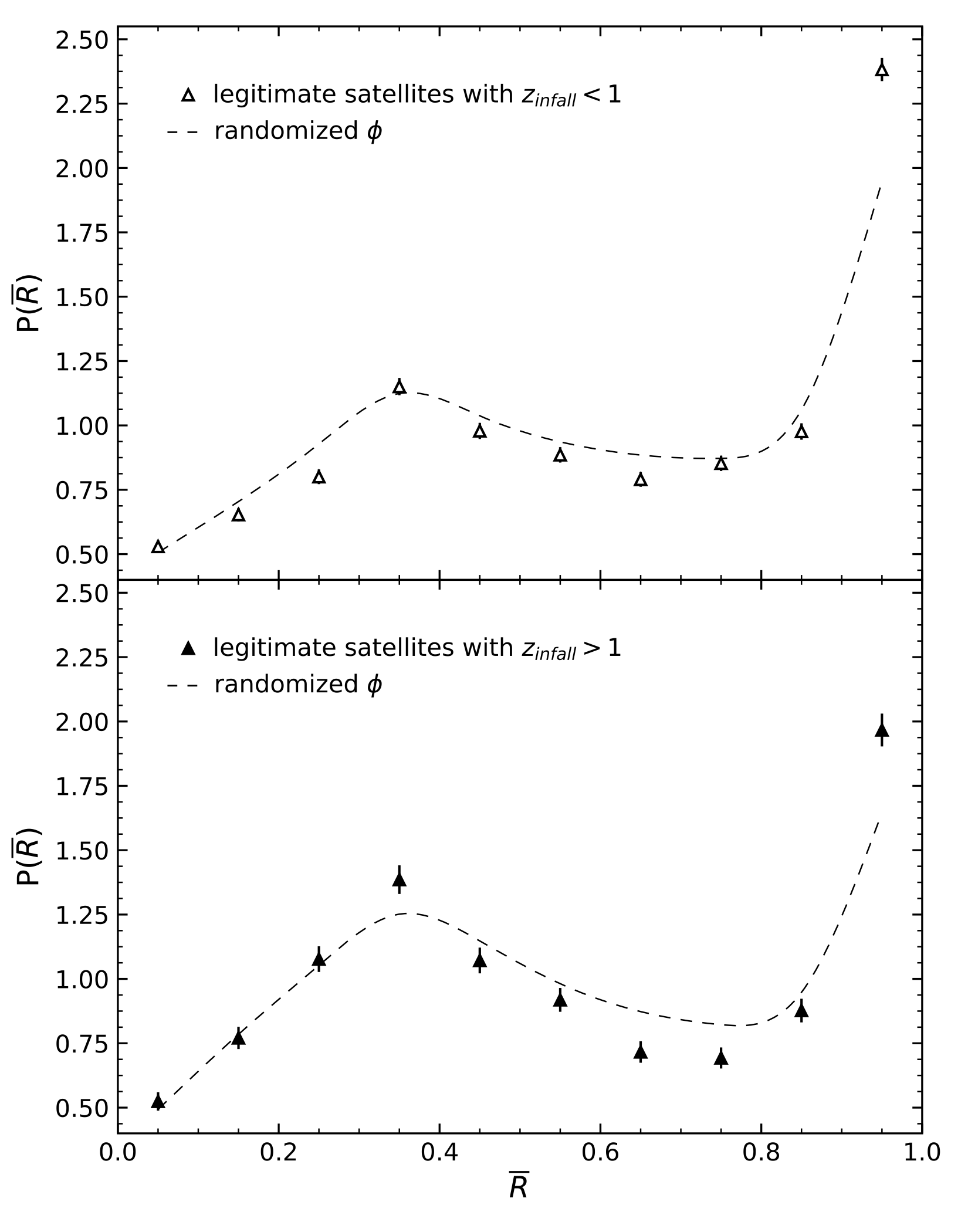}
    \caption{Probability distributions, $P(\overline{R})$.  Triangles: legitimate Millennium satellites.  Dashed lines: expected distributions for samples with random satellite locations. {\it Top:} satellites with $z_{\rm infall} < 1$.  {\it Bottom:} satellites  with $z_{\rm infall} > 1$.}
    \label{fig:p_of_rbar_infall}
\end{figure}

Figures \ref{fig:p_of_delta_phi_infall} and \ref{fig:p_of_rbar_infall} show the probability distributions, $P(\Delta \phi)$ and $P(\overline{R})$, for legitimate Millennium satellites, separated into samples with $z_{\rm infall} < 1$ and $z_{\rm infall} > 1$. In all panels in Figures~\ref{fig:p_of_delta_phi_infall} and \ref{fig:p_of_rbar_infall}, the spatial distributions are clearly lopsided, and $\chi^2$ tests indicate that both $P(\Delta \phi)$ and $P(\overline{R})$ for satellites with $z_{\rm infall} > 1$ and $z_{\rm infall} < 1$ are inconsistent with random satellite distributions at confidence levels $> 99.999\%$.

Satellites with $z_{\rm infall} > 1$ have been inside their host's dark matter halo for $\gtrsim 7.8$~Gyr and might be expected to be a virialized population.  That is, since the dynamical time for a spherical halo is $\tau_{\rm dyn} \sim {0.1/H_0}$ (e.g., \citealt{2010RAA....10.1242G}), satellites with $z_{\rm infall} > 1$ have been within their host's halos for a time that is $\gtrsim 5.5 \tau_{\rm dyn}$. However, from Figures \ref{fig:p_of_delta_phi_infall} and \ref{fig:p_of_rbar_infall}, these ``oldest'' satellites are not distributed symmetrically about their hosts.  Compared to satellites with $z_{\rm infall} < 1$ (lopsidedness fraction, $f_{\rm lop} = 1.26 \pm 0.03$), however, satellites with $z_{\rm infall} > 1$ exhibit somewhat less lopsidedness in their spatial distribution ($f_{\rm lop} = 1.11 \pm 0.03$), which is consistent with the spatial distribution of the oldest satellites having had more time to be influenced by the gravitational potentials of their hosts' halos.


\section{Summary and Discussion} \label{sec:Summary_and_Discussion}

We have investigated the spatial distributions of the satellites of bright, isolated host galaxies in a $\Lambda$CDM universe.  The host-satellite sample was selected from mock redshift space lightcones (built upon the Millennium simulation, and updated for the Planck 2014 cosmology) using the criteria adopted by BS20 in our study of the spatial distributions of observed satellite galaxies in the NASA-Sloan Atlas (NSA).  Obtaining the theoretical host-satellite sample in the same way as the observational host-satellite sample allows the most direct comparison of the observed universe to the predictions of $\Lambda$CDM, and also allows potentially spurious sources of lopsidedness in the observed satellite distribution to be assessed.

From Table~\ref{table:sample_properties} and Figure~\ref{fig:sample_properties}, the physical properties of the $\Lambda$CDM and observational host-satellite samples are quite similar overall.  The most notable differences between the two samples are that Millennium sample is somewhat bluer than the NSA sample, and the mean Millennium host-satellite luminosity ratio is somewhat higher than the mean NSA host-satellite luminosity ratio.

We used two metrics to quantify the spatial distributions of the satellite galaxies, relative to their hosts: the pairwise polar angle difference, $P(\Delta\phi)$, and the mean resultant length, $\overline{R}$.  Both metrics indicate that the spatial distributions of the Millennium satellites are lopsided, with a strong tendency for satellites to be found preferentially on one side of the host (in the plane of the sky).  In agreement with the observational results of BS20, the lopsidedness is greatest for Millennium systems with blue hosts and Millennium systems with 2 to 3 satellites. 

The pairwise polar angle difference characterizes the clustering of {\it pairs} of satellites, aggregated over all host-satellite systems.  In contrast, the mean resultant length characterizes the ``directionality'' (relative to the host) of the satellites within each {\it individual system}.  These are complementary metrics and comparison of our results for the Millennium and NSA samples leads to different conclusions regarding the degree to which the results from the two samples agree with each other.  We find that the degree of clustering of pairs of satellites around blue host galaxies and pairs of satellites in systems with only 2 to 3 satellites is essentially the same in both the Millennium and NSA samples.  However, the degree of clustering of pairs of satellites in all other cases (complete sample, satellites of red hosts, systems with 4 to 6 satellites, systems with 7 to 21 satellites) is much greater in the Millennium sample than it is in the NSA sample.  When we examine the directionality of the satellites within each system, we find that, with the exception of the largest systems (7 to 21 satellites), the overall directionality of the satellite distributions in the Millennium sample is consistent with the overall directionality of the satellite distributions in the NSA sample.  In the case of the largest host-satellite systems, the directionality of the satellite distribution is somewhat greater for the Millennium sample than it is for the NSA sample.

Two obvious sources of potentially spurious lopsidedness in the distribution of NSA satellites are: [1] satellites that are missing from the NSA sample because their spectra do not exist in the source databases (i.e., incompleteness of the satellite catalog itself) and [2] interloper galaxies (i.e., galaxies that pass the satellite selection criteria but are not physically associated with a host galaxy).  Since the redshifts of all Millennium galaxies are known, and since there is a pronounced lopsidedness to the distribution of Millennium satellites, it is highly unlikely that the lopsidedness shown by the NSA satellites is solely attributable to incompleteness of the satellite catalog.  Because the 3D distribution of all Millennium galaxies is known, it is straightforward to determine which objects in the Millennium satellite sample are outside the virial radii of the hosts.  When we eliminate these objects from our analysis, the distribution of the remaining (``legitimate'') satellites is still distinctly lopsided.  We find that, while the presence of interlopers increases the lopsidedness of the Millennium satellite distribution, the lopsidedness of the Millennium satellite distribution is by no means solely attributable to the interlopers.  Therefore, it is unlikely that the lopsidedness of the NSA satellite distribution is solely attributable to interlopers.
Since satellites that have been within their hosts' dark matter halos for several dynamical times can be expect to be a virialized population, it is also possible that the lopsidedness of the satellite distribution is solely attributable to satellites that have only recently arrived within their hosts' halos.  However, we find that this is not the case for the Millennium satellites.  While satellites that entered their hosts' halos most recently ($z_{\rm infall} < 1$) do show a greater degree of lopsideness in their spatial distribution, satellites that have been within their hosts' halos for $\sim 8$~Gyr or more also have spatial distributions that are significantly lopsided.
In agreement with W21, who obtained isolated host galaxies and their satellites from the $\Lambda$CDM TNG300 simulation using 3D selection criteria and whose host galaxies were typically more massive than ours (i.e., $\sim 70$\% of the Millennium host galaxies have stellar masses that are less than the minimum host stellar mass adopted by W21), we find that the lopsidedness of the Millennium satellite distributions depends upon the color of the host galaxy, with the satellites of blue host galaxies having a greater degree of lopsidedness in their spatial distribution than the satellites of red host galaxies.  Also, while the host samples are entirely different (i.e., isolated hosts vs.\ pairs of hosts), it is notable that even the oldest ``legitimate'' Millennium satellites have significantly lopsided spatial distributions, which contrasts strongly with the results of \citet{2019MNRAS.488.3100G} who found that the lopsided satellite distribution for pairs of large galaxies is attributable to satellites that have not been present long enough to have experienced a ``fly-by''.  

\section{Conclusions} \label{conclusions}

The main conclusions from this work are the following.  When samples of isolated host galaxies and their satellites are obtained from a $\Lambda$CDM simulation using the same type of criteria that are used to obtain host-satellite samples from a redshift survey, the spatial distribution of the satellites shows a pronounced degree of lopsidedness.  In agreement with observations, the lopsidedness is driven by the satellites of blue host galaxies and satellites in systems with relatively few (2 to 3) satellites.  The lopsidedness found in the observational samples is not solely attributable to incompleteness in the satellite sample.  In addition, while the presence of interlopers (i.e., false satellites) accounts for some degree of lopsidedness in the satellite distribution, the lopsidedness is not solely attributable to interlopers.  Furthermore, the lopsidedness is not solely attributable to satellites that have only recently arrived within their hosts' halos.

\begin{acknowledgments}
This work was supported by National Science 
Foundation grant AST-2009397.
\end{acknowledgments}

\bibliography{bib_file_2021}{}
\bibliographystyle{aasjournal}

\end{document}